\documentclass[prb,twocolumn,floatfix,preprintnumbers,amsmath,amssymb,superscriptaddress]{revtex4}
\usepackage{graphicx}
\usepackage{dcolumn}
\usepackage{bm}
\usepackage{graphicx}
\usepackage{color}
\usepackage{bm}

\newcommand{\pfn}{PbFe$_{1/2}$Nb$_{1/2}$O$_{3}$\,}
\newcommand{\pfns}{PFN\,}

\definecolor{reddeep}{rgb}{0.995,0.000,0.200}

\begin{document}
\title
{Microscopic Coexistence of Antiferromagnetic and Spin-Glass States}

\author{S. Chillal}
\affiliation{Neutron Scattering and Magnetism, Laboratory for Solid State Physics, ETH Z\"urich, Z\"urich, Switzerland}
\author{M. Thede}
\affiliation{Neutron Scattering and Magnetism, Laboratory for Solid State Physics, ETH Z\"urich, Z\"urich, Switzerland}
\affiliation{Laboratory for Muon Spin Spectroscopy, Paul Scherrer Insitut, Villigen-PSI, Switzerland}
\author{F.J. Litterst}
\affiliation{Institut f\"{u}r Physik der Kondensierten Materie, Technische Universit\"at Braunschweig, 38106 Braunschweig, Germany}
\author{S.N. Gvasaliya}
\email[]{sgvasali@phys.ethz.ch}

\affiliation{Neutron Scattering and Magnetism, Laboratory for Solid State Physics, ETH Z\"urich, Z\"urich, Switzerland}
\author{T.A. Shaplygina}
\affiliation{Ioffe Physico-Technical Institute RAS, 194021, St. Petersburg, Russia}
\author{S.G. Lushnikov}
\affiliation{Ioffe Physico-Technical Institute RAS, 194021, St. Petersburg, Russia}
\author{A. Zheludev}
\email[]{zhelud@ethz.ch}
\homepage[]{http://www.neutron.ethz.ch/}
\affiliation{Neutron Scattering and Magnetism, Laboratory for Solid State Physics, ETH Z\"urich, Z\"urich, Switzerland}

\begin{abstract}
The disordered antiferromagnet \pfn (\pfns) is investigated in a wide temperature range by combining M\"ossbauer spectroscopy and neutron diffraction experiments.
It is demonstrated that the magnetic ground state is a {\it microscopic} coexistence of antiferromagnetic and a spin-glass orders.
This speromagnet-like phase features frozen-in short-range fluctuations of the Fe$^{3+}$ magnetic moments that are transverse to the long-range ordered
antiferromagnetic spin component.
\end{abstract}

\date{\today}

\maketitle

Phase transitions in the presence of disorder and/or competing interactions are one of the central unresolved problems in modern condensed matter physics~\cite{dagotto2005,balents,zheniya2008,anderson}. With both effects present, one may encounter a freezing of microscopic degrees of freedom without conventional long-range order. In magnetic systems, the corresponding phenomenon is referred to as a spin-glass (SG) transition~\cite{binder1986}. By now, spin glasses are reasonably well understood for models with discrete (Ising) symmetries and long-range interactions~\cite{ballesteros,palassini}. In contrast, for continuous (Heisenberg and XY) symmetries with short-range coupling, the properties and sometimes the very existence of the SG phase remain a matter of debate~\cite{abiko,lee2003,viet2009,viet2010,young2008}. An important outstanding question is whether the SG phase can coexist with true long-range order (LRO)~\cite{young2008,binderbook}? Theory~\cite{sk,gt,shender1987} and numerical studies~\cite{matsubara1992,ryan1996,matsubara2005,ryan_2012} have consistently provide an affirmative answer; see Ref.~\cite{ryan1992r} for a review. Both ferromagnetic (FM)~\cite{gt} and AF ~\cite{shender1987} models demonstrate a SG freezing of spin components transverse to the long range order parameter. The problem gained a particular urgency in the context of cuprate superconductors, where SG and AF phases are adjacent on the concentration-temperature phase diagram but appear to be mutually exclusive~\cite{niedermayer,alloul2009}.

On the experimental side though, the situation is much less clear-cut and hotly debated. Most hurdles on this route are the known measurement issues endemic to spin glasses~\cite{dagotto2005,dagotto2001,johnston2010}. In addition, even if long range order and SG are shown to appear simultaneously, it may be extremely difficult to establish their co-existence on the {\it microscopic} scale, as opposed to an inhomogeneous phase separation.
A great deal of work was done on {\it amorphous, ferromagnetic} Fe$_{X}$Zr$_{100-X}$ alloys. While strong support for uniformly coexisting SG
and LRO in these systems have been presented~\cite{ryan1995,ryan_1987,ryan_2000,ryan_2012}, evidence pointing to a cluster-based scenario also exist~\cite{kaul1992}. In crystalline materials, simultaneous antiferromagnetic (AF) and SG states have been observed in  Fe$_{0.6}$Mn$_{0.4}$TiO$_3$~\cite{ito1987,ito1996} and Co$_2$(OH)PO$_4$~\cite{rojo_2002}. However, even in these {\it Ising} systems, the microscopic nature of such coexistence is not unambiguous~\cite{ito1996,rojo_2002}.

A solid experimental proof of microscopic SG and LRO coexistence in a crystalline material remains elusive. Besides finding an appropriate model compound, one has to strategically choose the experimental techniques. Momentum-resolved (scattering) experiments are well-suited to probe microscopic quantities averaged over the entire sample, but do not provide spatially-resolved information. In contrast, local-probe resonant methods are ideal tools for validating  homogeneity on the microscopic level, but carry no information on spatial coherence.
Only by merging the two approaches the coexistence of complex phases can be unambiguously established. In this Letter we report combined M\"ossbauer spectroscopy and neutron scattering experiments on the disordered AF \pfn. We prove that in this {\it crystalline Heisenberg AF system} SG and true AF long-range order coexist on the {\it microscopic} scale.

\begin{figure}
\includegraphics[width=0.9 \columnwidth]{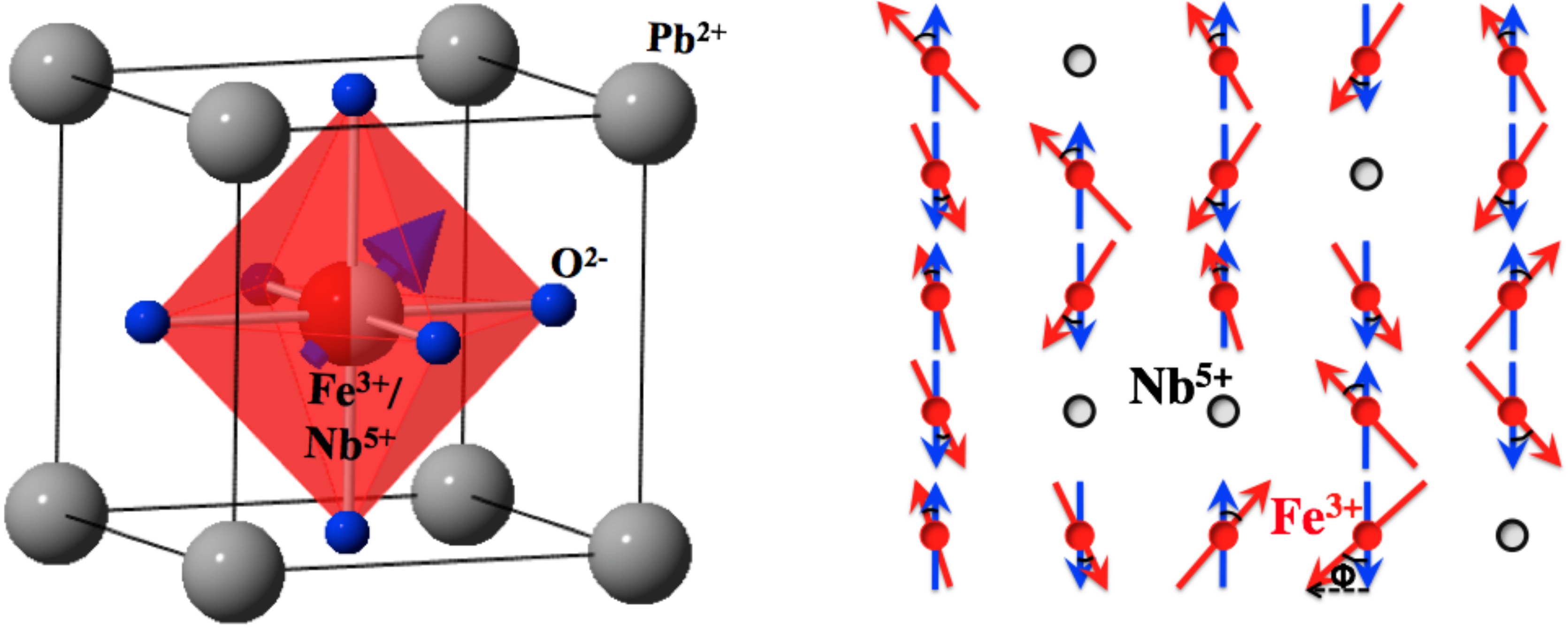}
\caption{Left: Structural motif of \pfn. Right: Antiferromagnetic G-type order of Fe$^{3+}$ moments in the collinear phase (blue arrows) and proposed model for speromagnetic-like ordering in the antiferromagnetic spin glass phase (red arrows). The tilt angle $\Phi$ randomly varies from site to site averaging to zero.  Nb$^{5+}$ are non-magnetic.}
  \label{fig:ss}
  \end{figure}

Our target material, abbreviated \pfns, is a well-known complex perovskite. Its structure is shown in Fig.~\ref{fig:ss} (left panel). The Pb ions reside at the corners of the unit cell, while oxygen octahedra surrounds the Fe and Nb sites. The material is nearly cubic with a lattice constant $a= 4.01$~\AA~\cite{bonny1997}. Although the spin Hamiltonian of PFN has not been established, it is believed that nearest neighbor Heisenberg interactions are frustrated by next nearest neighbor AF ones. The magnetic Fe$^{3+}$ and Nb$^{5+}$ are randomly distributed over the B-sites of the perovskite lattice~\cite{darlington1991,bonny1997}. The non-magnetic sites locally relieve the geometric frustration of interactions in a spatially random manner. AF long-range order occurs below T$\rm_N\sim145$~K. The Fe$^{3+}$ moments are arranged in a simple $G$-type structure~\cite{ivanov} as shown by blue arrows in the right panel of Fig.~\ref{fig:ss}. The spin glass state emerges at a lower temperature, $T_\mathrm{SG}\sim12$~K. It is manifested in a difference between magnetization curves measured in zero-field-cooled (ZFC) and field-cooled samples~\cite{falqui}. What is known from combined muon-spin rotation ($\mu$SR) and neutron scattering studies, is that the long-range AF order in \pfns is not destroyed by the appearance of SG state~\cite{gelu2009}. However, based on the magnetoelectric (ME) experiments, it has been argued that this coexistence is a phase separation~\cite{kleemann2010}. In this model, the two type of order emerge independently on separate subsystems: infinite-range percolation cluster (AF) and isolated Fe$^{3+}$ ions and unblocked superantiferromagnetic Fe$^{3+}$ clusters (SG). Below we shall demonstrate that, in fact, the coexistence is a single homogeneous phase, with  the magnetic moments arranged in a speromagnetic-like fashion, as indicated by the red arrows in the right panel of Fig.~\ref{fig:ss}b.

\begin{figure}
\includegraphics[width=0.9 \columnwidth]{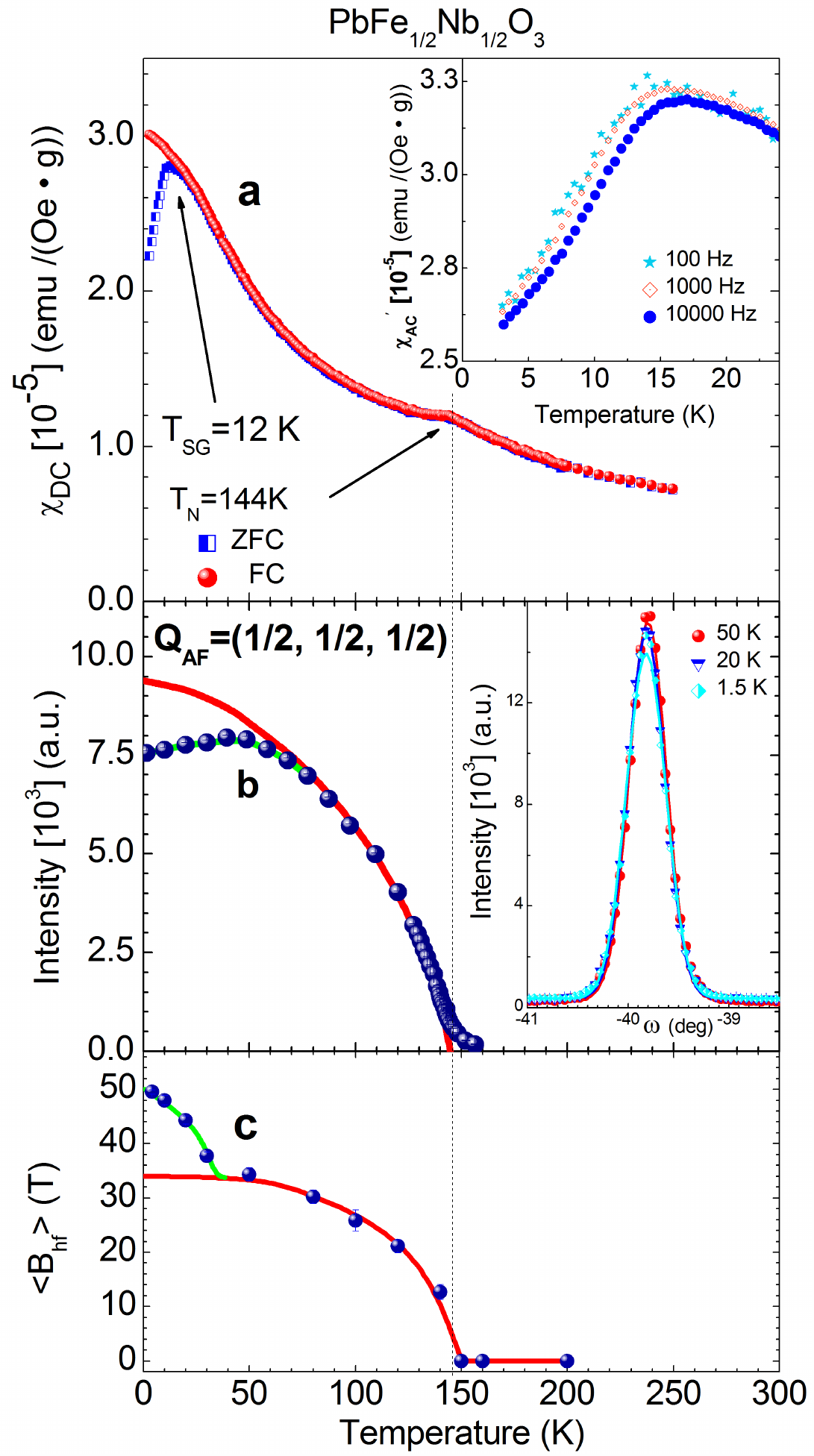}
\caption{(a) DC susceptibility of \pfns taken measured in 100 Oe applied field following ZFC and FC protocols. Inset: AC-susceptibility taken in the range 10$^2$ - 10$^4$~Hz (b) Measured temperature dependence of the
intensity of an antiferromagnetic Bragg peak in \pfns (symbols) \protect\footnote{It is not feasible reliably calculate the Brillouin curve for PFN, as the temperature dependence of the staggered magnetization is not a
monotonous function, and the ordered moment is much reduced from the expectations for Fe$^{3+}$~\cite{ivanov}}. The evolution of an AF Bragg peak for T$\le 50$ is shown in inset. (c) Average hyperfine field $\langle B_{hf}\rangle$ obtained from the M\"ossbauer spectra as a function of temperature. All lines are guides for the eye. }
\label{fig:tdep}
\end{figure}

In our \pfns samples~\cite{samples}, the two phase transitions are readily observed in macroscopic experiments. Figure~\ref{fig:tdep}a shows magnetic DC susceptibility versus temperature obtained using ZFC and FC protocols~\cite{macro}. Below $T_\mathrm{SG}\sim 12$ K the two curves diverge as is typical for a spin glass. SG behavior is further evidenced by the gradual frequency dependence of the rounded peak in AC susceptibility (Figure~\ref{fig:tdep}a, inset). In contrast, the small cusp in susceptibility which is due to AF ordering at $T_\mathrm{N}$ does not show noticeable history effects (see further details in the Supplement). These observations are in agreement with previous reports~\cite{falqui,kleemann2010,kumar2008}, confirming that our samples are essentially identical to those used by other groups. Of course, as any true long range ordering, the AF phase transition is best represented by the emergence of a new AF Bragg peak in neutron diffraction. The measured temperature dependence of the ${\mathbf Q}=(1/2,1/2,1/2)$ Bragg intensity is plotted in Figure~\ref{fig:tdep}b~\cite{neutron}. As expected for an AF, the intensity of the magnetic Bragg peak ({\it i.e.} square of magnetization) increases smoothly below T$_{\rm N}$. Below T$_\mathrm{SG}$, a small decrease of the AF Bragg peak intensity is observed, whereas the lineshape remains Gaussian.
This observation is reminiscent of that seen in certain reentrant systems~\cite{mirebeau1991,ito1987}, and may be attributed to spin-canting. As discussed below, this is precisely our interpretation of this effect in the case of \pfns.

Verifying the coexistence of the SG and AF phases  the microscopic level calls for the use of microscopic local probes. For this purpose M\"ossbauer spectroscopy employs magnetic nuclei already present in the material, in our case those of $^{57}$Fe. In a paramagnetic state, the position of the nuclear absorption line is the isomer (chemical) shift, related to the valence state of the ion. This line may exhibit additional quadrupolar splitting due to a non-spherical charge distribution around the  ion, resulting in an electric field gradient at the nuclear site. In magnetically ordered phases, the degeneracy of the nuclear energy levels is further lifted by the local hyperfine field at the nuclei, generated by the static sublattice magnetization. The resulting energy intervals between the lines are then a measure of the
iron magnetic moment.

\begin{figure}
\includegraphics[width=0.9 \columnwidth]{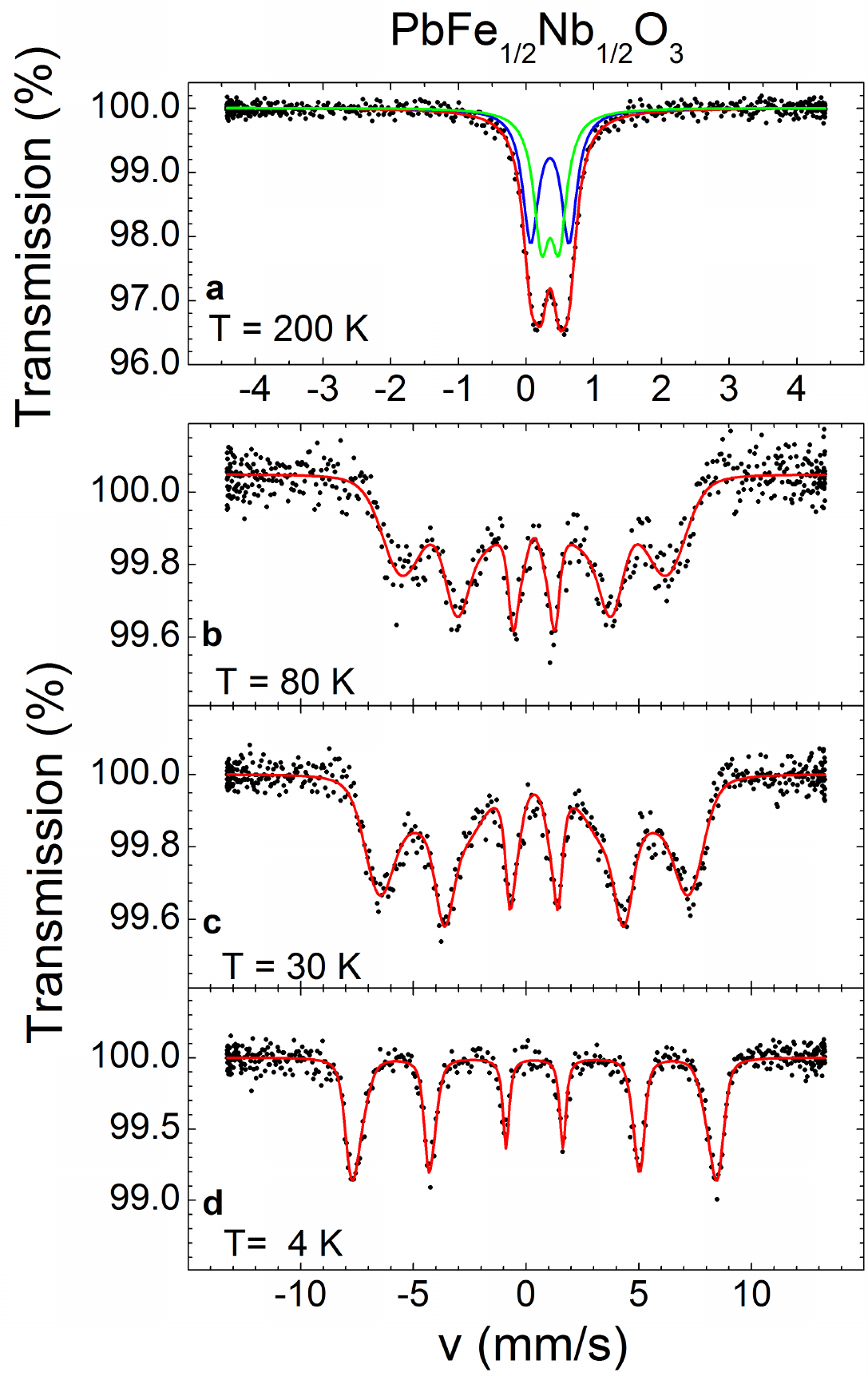}
\caption{Representative M\"{o}ssbauer spectra of \pfns taken in the paramagnetic phase (a) and below its N\'eel temperature (b-d). Note
a reduced velocity scale in panel (a).}
  \label{fig:spectra}
\end{figure}

The way this scenario plays out in \pfn  is illustrated by the typical  M\"ossbauer spectra shown in Fig.~\ref{fig:spectra}~\cite{mossb}. All data were collected on powder samples prepared by grinding either single crystals
or ceramic material, with virtually identical results. This consistency proves that all effects described below are robust and sample-independent.At T=200~K, in the paramagnetic phase, a single structured line is
observed (Fig.~\ref{fig:spectra}a).
The additional slight splitting is due to the above-mentioned quadrupolar effect ~\cite{raevski2012}. For all samples studied in this work, the spectra measured at $T>T_\mathrm{N}$ can be fit (red line in
Fig.~\ref{fig:spectra}) assuming two types of Fe$^{3+}$ sites with distinct isomer shifts (blue and green line). Attempts to fit the paramagnetic spectra with a single contribution gave less satisfactory agreement factors
and a significant broadening. The two effects become progressively more pronounced in the AF phase, suggesting that the one-component model is inappropriate. As discussed elsewhere~\cite{raevski2012}, the two different sites
may correspond to local variations in chemical short range order. Our analysis yields approximately equal populations for the two components~\cite{ratio} . The obtained isomer shifts can be unambiguously attributed to
trivalent high spin iron Fe$^{3+}$.

\begin{figure}
\includegraphics[width=0.9 \columnwidth]{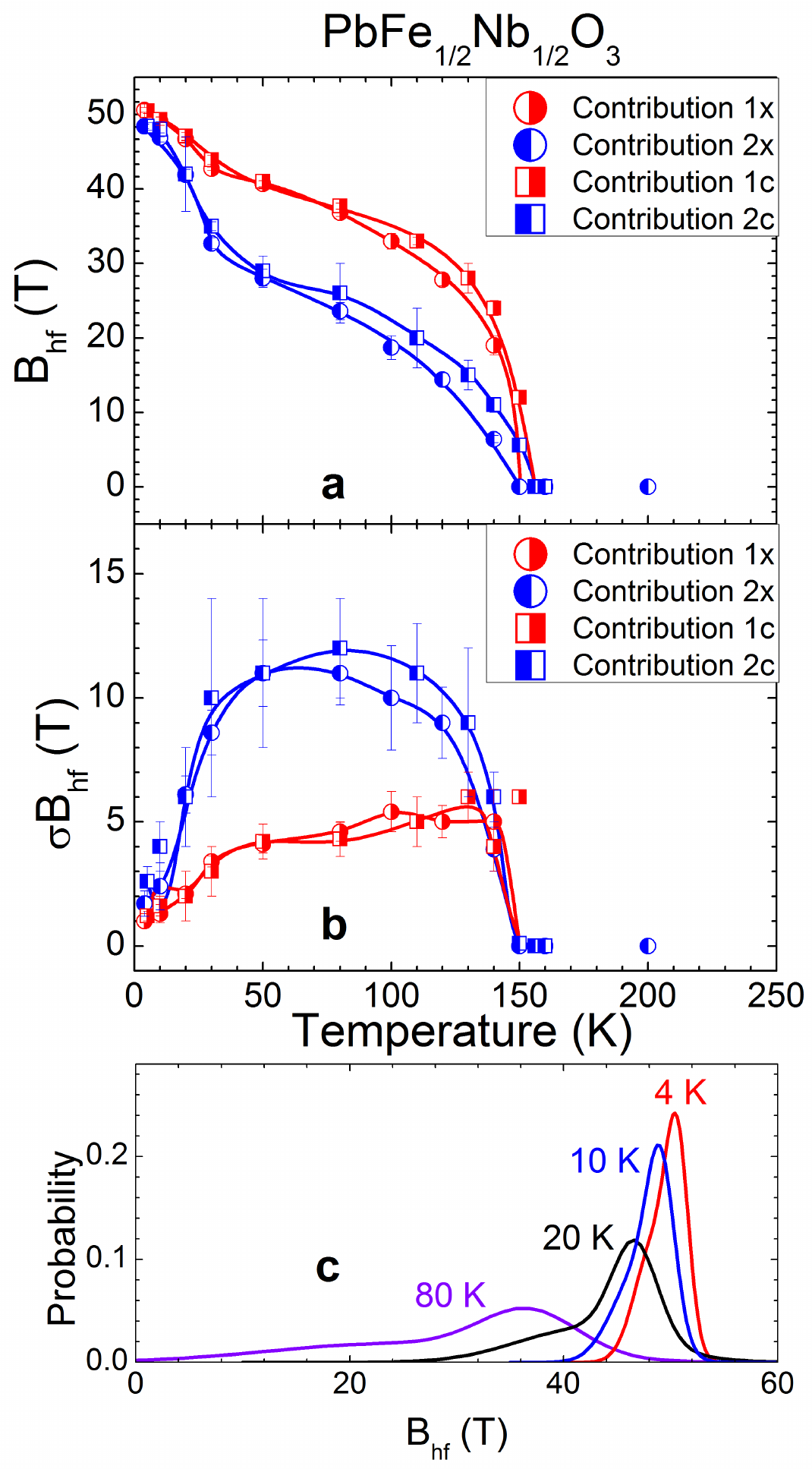}
\caption{ (a, b)The magnetic hyperfine fields for the two Fe$^{3+}$ sites (1,2) as deduced from the measured M\"ossabuer spectra, as described in the text. Labels
         ``c'' and ``x'' stand for ceramic and crystalline samples, respectively. (c)  Evolution of the P(B$_{hf}$) towards homogeneous ground state.}
  \label{fig:4}
\end{figure}

Below $T_\mathrm{N}$, the single line is split due to the appearance of long range order, as shown in Figs.~\ref{fig:spectra}~b--d. The temperature dependence of the measured average hyperfine field $\langle B_{hf} \rangle$
is presented in Fig.~\ref{fig:tdep}c. In the AF phase, it follows the magnetic order parameter, as expected. However, a crucial new observation is that the unsplit (paramagnetic) component that dominated in the paramagnetic
state, is no longer present below $T_\mathrm{N}$. This immediately rules out the phase separation model of Ref.~\cite{kleemann2010}. In the latter, the low temperature SG phase is confined to isolated spin clusters.
Above $T_\mathrm{SG}$ these would simply be paramagnetic, and necessarily produce a resonance line almost identical to that above $T_\mathrm{N}$. In contrast, our data show that {\it all spins in the system are involved in
the AF order}. At the same time we note that the quadrupolar splitting, so clearly visible at high temperatures, is absent below $T_\mathrm{N}$. This indicates a wide distribution of angles between the iron spin and the local
axes of the electric field gradient tensor.

Important clues to the nature of the ordered phases are obtained by a more thorough quantitative analysis~\cite{raevski2012,Yang}.
The simplest strategy is to treat the spectra as superpositions of two contributions with distinct hyperfine fields.
Even so, to obtain a good fit, one has to allow for an intrinsic and temperature-dependent broadening of the absorption lines, which implies a {\it distribution of magnetic hyperfine fields}. Assuming these distributions
P(B$_{hf}$) to be Gaussian, and  fitting the data to the resulting model, produces the fits shown in red lines in  Figs.~\ref{fig:spectra}~b--d. The procedure consistently yields equal spectral weights of the
two components at all temperatures. The hyperfine fields for each one are separately plotted as a function of temperature in Fig.~\ref{fig:4}a.

Another experimental result that firmly supports a microscopic coexistence of AF and SG orders is the measured temperature dependence of the width  $\sigma B_{hf}$ of the hyperfine field distribution (Fig.~\ref{fig:4}b).
For both sites, $\sigma B_{hf}$ gradually increases on cooling below T$\rm_N$ in the AF phase and reaches a broad maximum between 120~K and 50~K. Below $T\sim 50$~K, the distribution width decreases drastically, and remains
very small and constant below $T_\mathrm{SG}\sim$12K. The most natural interpretation of such behavior is that line broadening is due to slow fluctuations of local magnetic fields within M\"{o}ssbauer frequency window
(MHz to GHz). In this scenario, all of the available Fe spins are involved in creating the AF order, and the simultaneous narrowing of all lines at low temperatures corresponds to all of them being involved in SG
freezing. Such behavior is totally inconsistent with the above-mentioned inhomogeneous model. Indeed, a partition into smaller dynamic clusters and infinitely connected static AF cluster would result in narrower lines in
the high temperature AF phase.

Our conclusion is further supported by the temperature evolution of the distribution of magnetic hyperfine fields, which is known to provide valuable clues on the spin
arrangement~\cite{morop2003,bocquet1992,alex2012,kaul1992,ryan1995}. The distributions determined for \pfns are shown in Fig.~\ref{fig:4}c and further details are given in the Supplement.
Whereas above ~$\sim10$~K it exhibits pronounced double-peak shape, just
a narrow peak with a slight shoulder is observed at base temperature. These findings are precisely opposite to the expectations for inhomogeneous (cluster-like) scenario~\cite{kaul1992}.

That the magnetic state below about 10~K is homogeneous as sensed by the M\"{o}ssbauer nuclei, is also supported by the nearly coinciding saturation values
of $B_{hf}$ for the two contributions.  These value are close to an average hyperfine field at saturation typically found in other Fe$^{3+}$-based
perovskites~\cite{shirane1962}, suggesting a full recovery of the Fe$\rm^{3+}$ moment at the lowest temperature.

We are now in a position to propose a model for the low-temperature magnetic structure of \pfns. As mentioned, the variation of $B_{hf}$ above 50~K (Fig.~\ref{fig:tdep}c) roughly follows that of the AF long range order
parameter determined by neutron diffraction (Fig.~\ref{fig:tdep}b). It stands to reason that in this regime it mainly reflects the evolution of the collinear long range ordered AF component $\langle S^{z}(T)\rangle$
of Fe$\rm^{3+}$ (Fig.~\ref{fig:ss}b, blue arrows). The increase of mean magnetic hyperfine fields on cooling below
50~K (Fig.~\ref{fig:tdep}c, green line), is reminiscent of  precursor phenomena in ferromagnetic re-entrant spin-glass systems like AuFe~\cite{Lauer,BrandPRB} or the mixed spinel Mg$_{1+t}$Fe$_{2-2t}$Ti$_t$O$_4$~\cite{BrandFMP}.
It represents a gradual freezing of transverse spin components within M\"{o}ssbauer frequency window. In other words, there occurs a tilting of the ionic moments from the $z$ direction set by the AF long-range order
(Fig.~\ref{fig:ss}b, red arrows).
The process culminates in a complete (static) freezing at $T_\mathrm{SG}$. From the change of $B_{hf}$ observed between 50~K and 4~K, we can roughly estimate the typical tilt angles to be $\sim 30 \rm ^o$ and
$\sim 55\rm ^o$.

Unlike the longitudinal component, the transverse one show no long-range correlations. At the time-scale of $\sim10^{-11}$~s  it contributes to the static diffuse neutron magnetic scattering with a correlation
length $\sim15$~\AA~\cite{gelu2009}, but adds nothing to the AF Bragg peak intensity. In our proposed model, the progressively more pronounced transverse correlations cause a canting of spins below T$\sim$50~K, and become
liable for the reduction of the latter (Fig.~\ref{fig:tdep}b, green line).

In conclusion, at low temperatures, the crystalline Heisenberg system \pfn realizes a unique combination of antiferromagnetic long range order for one set of spin components, and a spin glass state for the transverse ones.

This work is partially supported by the Swiss National Fund through MANEP.

\bibliographystyle{prsty}

\begin{thebibliography}{99}
\bibitem{macro}Magnetization and susceptibility measurements were performed using a PPMS from Quantum Design.
\bibitem{mossb} Conventional $^{57}$Fe M\"{o}ssbauer absorption spectroscopy experiments in transmission geometry we performed on 60 mg/cm$^{2}$
powder samples, which corresponds to 0.1 mg/cm$^{2}$ of $^{57}$Fe. The absorbers were kept within nylon containers in a static helium exchange~gas
atmosphere of about 0.2 mbar in a liquid helium flow cryostat.
\end{thebibliography}

\end{document}